\documentclass[12pt,paper]{iopart}

\usepackage{graphicx}
\usepackage{dcolumn}
\usepackage{bm}
\usepackage{color}
\usepackage[urlcolor=blue,urlbordercolor={0 0 1}]{hyperref}

\begin{document}

\title{Identifying single electron charge sensor events using wavelet edge detection}

\author{J~R~Prance$^{1,2}$, B~J~Van~Bael$^1$, C~B~Simmons$^1$, D~E~Savage$^1$, M~G~Lagally$^1$, Mark~Friesen$^1$, S~N~Coppersmith$^1$, M~A~Eriksson$^1$}
\address{$^1$ University of Wisconsin-Madison, Wisconsin 53706, United States}
\address{$^2$ Department of Physics, Lancaster University, Lancaster, LA1 4YB, United Kingdom}

\begin{abstract}
The operation of solid-state qubits often relies on single-shot readout using a nanoelectronic charge sensor, and the detection of events in a noisy sensor signal is crucial for high fidelity readout of such qubits. The most common detection scheme, comparing the signal to a threshold value, is accurate at low noise levels but is not robust to low-frequency noise and signal drift. We describe an alternative method for identifying charge sensor events using wavelet edge detection. The technique is convenient to use and we show that, with realistic signals and a single tunable parameter, wavelet detection can outperform thresholding and is significantly more tolerant to $1/f$ and low-frequency noise.
\end{abstract}

\noindent{\it This is an author-created, un-copyedited version of an article accepted for publication/published in Nanotechnology. IOP Publishing Ltd is not responsible for any errors or omissions in this version of the manuscript or any version derived from it. The Version of Record is available online at \href{http://dx.doi.org/10.1088/0957-4484/26/21/215201}{doi:10.1088/0957-4484/26/21/215201}.}

\maketitle


\section{Introduction}

Charge sensors with single-electron sensitivity are essential to the operation of many solid-state qubits. These sensors can be realized using nanoelectronic devices such as single-electron transistors (SETs), quantum point-contacts (QPCs) and quantum dots, all of which are highly sensitive to their local electrostatic environment. The sensor can measure the state of a charge qubit directly \cite{Field1993,Elzerman2003,Petta2004,Astafiev2004,Shi2013} or, via spin-to-charge conversion, the state of a spin qubit \cite{Elzerman2004,Hanson2005,Johnson2005a,Petta2005,Shi2012,Kim2014}. If the sensor bandwidth is large enough, the qubit state may be detected on a timescale shorter than its lifetime. Such single-shot detection is essential for implementing error correction in a quantum information processing architecture \cite{Ladd2010}. It can also be used to observe correlations between qubits \cite{Nowack2011,Shulman2012} and to measure the dynamics of the environment \cite{Barthel2012,Fink2013}. Single-shot readout with a charge sensor has been realized for a variety of solid-state qubit candidates, such as semiconductor quantum dots in GaAs \cite{Lu2003,Fujisawa2004a,Vandersypen2004,Schleser2004,Elzerman2004,Hanson2005,Meunier2006,Gustavsson2006a,Reilly2007,Barthel2009,Gustavsson2009,Mason2010}, InAs nanowires \cite{Gustavsson2008a}, carbon nanotubes \cite{Gotz2008}, graphene \cite{Guttinger2011}, silicon \cite{Morello2010,Thalakulam2011,Simmons2011,Mahapatra2011,Prance2012,Yuan2012}, and superconducting charge qubits \cite{Schoelkopf1998,Astafiev2004}. In all cases the signal from the charge sensor takes one of two or more discrete levels, where each level corresponds to a different configuration of charge in the qubit.

Numerous sources of noise can pollute the charge sensor signal. Sources of white noise include shot noise and Johnson-Nyquist noise in warm parts of the measurement circuit. Amplifier noise, which is often dominant, may be approximated as white noise over limited frequency ranges \cite{Gustavsson2006a}. An intrinsic source of $1/f$ noise is charge fluctuations in the vicinity of the sensor and the qubit \cite{Jung2004}. Other potential noise sources include interference from AC power supplies, signal drift due to instability in the measurement electronics, and instability of the sensor itself. High fidelity readout of a qubit must therefore rely on the adoption of a filter or algorithm that can accurately identify real events in a noisy signal. Wavelet signal processing is one promising solution to this challenge.

Wavelet signal processing is a joint time-frequency analysis technique that is well-suited to identifying within a signal localised events that posses particular spectral characteristics. The technique has found numerous applictions ranging from genetics to image compression and storage of fingerprint data \cite{ENCODE2007, JPEG2000, Bradley1993}. One common use of wavelet analysis is the detection of sharp edges in images \cite{Mallat2009}. The problems of detecting edges in images and events in a charge-sensor signal are very similar: in both cases the signal of interest is a sharp step overlayed on a noisy background.

In this paper we assess the performance of a wavelet edge detection algorithm for identifying events in a single electron charge sensor signal using both experimental data and simulated signals. We compare wavelet detection to an approach based on a threshold signal value. Thresholding, and more sophisticated derivatives of this method, are commonly used to analyze charge sensor signals \cite{Astafiev2004,Elzerman2004,Hanson2005,Nowack2011,Shulman2012,Barthel2012,Schleser2004,Meunier2006,Barthel2009,Morello2010,Thalakulam2011,Simmons2011,Prance2012,Lambert2013}. Below we show that, while both techniques perform well at low noise, wavelet edge detection has much better tolerance to $1/f$ and low-frequency noise and is also more tolerant to white noise. The wavelet algorithm described below has previously been used to successfully analyze measurements of a Si/SiGe spin qubit with a QPC charge sensor \cite{Simmons2011}. The noise spectra used here to assess the detection techniques include examples that closely resemble those found in that work. Both techniques are tested over a range of noise profiles, and the results are applicable to any sensor where events are characterized by switches between discrete signal levels.

\section{Method}

\begin{figure}
\centering
\includegraphics{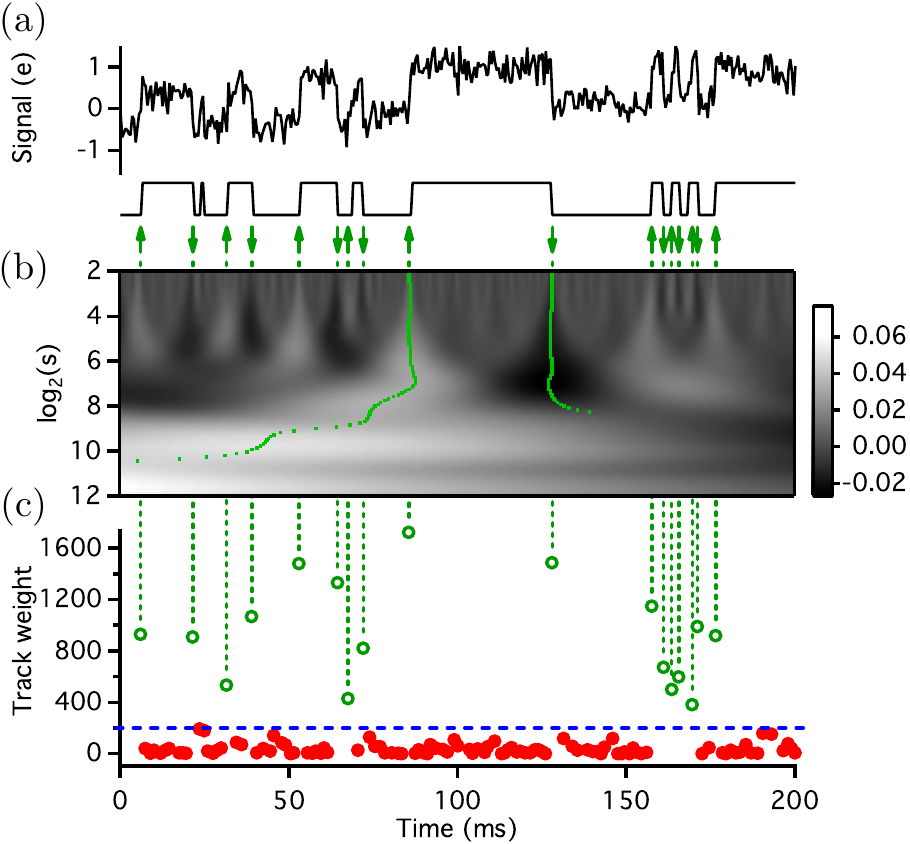}
\caption{Stages of wavelet edge detection. (a): first $200\,\mathrm{ms}$ of the input signal. The full signal has $2^{12}$ data points sampled at a rate of $2\,\mathrm{kHz}$. It was generated by adding simulated white and $1/f$ noise to the ``ideal'' signal shown below the axis. (b): first $200\,\mathrm{ms}$ of the multi-scale wavelet transform of the signal (see main text for details). Local maxima and minima in the transform are tracked in the direction of decreasing wavelet scale $s$. Two example tracks are shown by green lines. A total `weight' is calculated for each track (as described in the main text) and the results are shown in (c). Significant edges (open circles) are identified by tracks with a weight above $200$ (the dashed line). Other edges (closed circles) are ignored. The position of an edge is the position of its track at the finest scale ($s=2^{2}$). The edge direction (up or down) is given by the sign of the transform.\label{fig1}}
\end{figure}

Figure~\ref{fig1} outlines the stages of the wavelet edge detection process. The approach is both a formalization and generalization of Canny's edge detection algorithm \cite{Canny1986}. The detection is based upon a multi-scale real wavelet transform of the input signal. The transform $W$ at a time $t$ and wavelet scale $s$ is found by convolving the signal $I(t)$ with the scaled mother wavelet $\psi$:
$W(t, s) = \int_{-\infty}^{+\infty} I(t^\prime) \frac{1}{\sqrt{s}} \psi \left(\frac{t^\prime - t}{s}\right) dt^\prime\,$ \cite{Mallat2009}. By analogy with the Canny edge detection algorithm, we chose the mother wavelet $\psi$ to be the first derivative of a Gaussian. The scale $s$ sets the width of this wavelet and in Fig~\ref{fig1} and Fig~\ref{fig2} we express $s$ as a number of data points. An upwards (downwards) step in the signal will produce a local maximum (minimum) in the wavelet transform, and significant edges result in large, local maxima and minima that persist across a broad range of wavelet scales \cite{Mallat2009}.

Given a wavelet transform, the edge detection algorithm tracks local maxima and minima across wavelet scales to calculate a cumulative `weight' for every maximum/minimum present at the smallest scale. The weight of a point on a track is defined as the square of the transform normalised by the median value of the square of the transform at that scale. Significant edges in the signal are identified by tracks with a total weight above a cutoff value, and this is the only parameter used to optimize the detection. Our implementation of the algorithm is realized in MATLAB and the wavelet transformations are performed using the WaveLab850 library \cite{WaveLab2013}. 

Both experimental data and simulated charge sensor signals are used to investigate detection performance. The simulated signals are chosen to mimic typical measurements of Si/SiGe spin qubits using a QPC charge sensor \cite{Simmons2011,Prance2012}, and for easy comparison with typical data we present our results with a specific choice of time-scales on the abscissa; however, the results remain valid if either time and/or amplitude are scaled. The signals are generated by adding noise to an ideal charge sensor output. We choose a sampling rate of $2\,\mathrm{kHz}$ and a length of $2^{12}$ data points for all signals. The ideal output has two discrete levels at $\pm 0.5e$, where $e$ is the electron charge. Switches between the levels are generated randomly according to a Poisson distribution with a characteristic rate of $100\,\mathrm{Hz}$. A $1\,\mathrm{kHz}$ low-pass filter is applied to the ideal output, resulting in a signal rise-time that is common in experiments. After filtering, simulated white noise and $1/f$ noise are added to the signal.

The accuracy of the edge detection, when applied to a given signal, is measured by the F-score of the results $F = 2pr/(p + r)$, where the precision $p$ is the fraction of detected events that correspond to real events, and the recall $r$ is the fraction of events in the signal that were correctly detected \cite{Rijsbergen1979}. When calculating $p$ and $r$, we allow the maximum difference in time between real and detected events to be twice the time-constant of the low-pass filter ($2\,\mathrm{ms}$). $F=1$ indicates perfect detection. Both false negatives and false positives reduce $F$.

\section{Results and discussion}

We investigate the performance of wavelet edge detection using both simulated signals and experimental data. By using artificial signals, the true position of every event in the signal is known and the detection accuracy can be calculated. This analysis is presented in section~\ref{sec:simulated_signals}. In section~\ref{sec:real_signals} we investigate detection performance using measurements of a quantum point-contact charge sensor adjacent to a Si/SiGe double quantum dot.  The measured signals are known to have characteristics that enable the performance of the edge detection to be estimated, without knowing the true positions of the charging events.

\subsection{Comparison between wavelet and threshold detection using simulated signals}\label{sec:simulated_signals}

We compare wavelet edge detection to a simple alternative thresholding algorithm in which an event is detected when the signal crosses a threshold. We optimize both detection techniques for each signal by varying a single parameter to maximize $F$. In the case of threshold edge detection, this parameter is the threshold level. In the case of wavelet edge detection, the parameter is the final cutoff that determines whether a track in the transform has a large enough weight to be accepted as an event.

\begin{figure*}
\includegraphics[width=16cm]{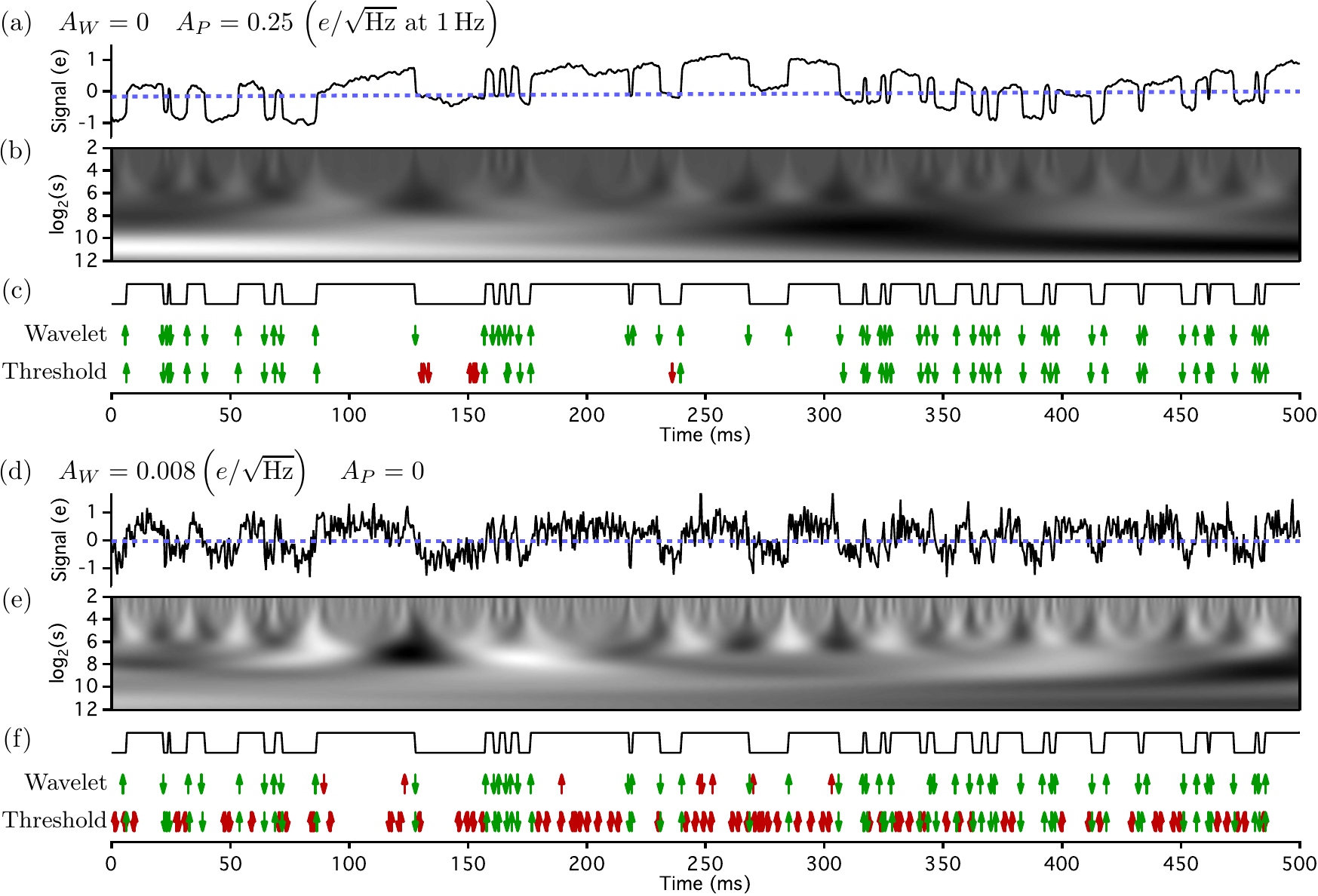}
\caption{Comparative examples of wavelet and threshold edge detection. (a-c) Results for a signal with $1/f$ noise of RMS amplitude $A_P$. (d-e) Results for a signal with white noise of RMS amplitude $A_W$. (a) and (d) show the input signals. The optimum levels for threshold edge detection are shown by blue dashed lines. (b) and (e) show wavelet transforms of the signals. (c) and (f) show the location of events given by wavelet and threshold edge detection, below the ideal sensor output. Correctly detected events are marked by green arrows. Detected events that do not correspond to a real event are marked by red arrows.
\label{fig2}}
\end{figure*}

Figure~\ref{fig2} shows the results of applying wavelet edge detection and the thresholding technique to two example signals. The signal in Fig.~\ref{fig2}(a) was generated with $1/f$ noise of RMS amplitude of $A_P = 0.25e\,\mathrm{/ \sqrt{Hz}}\,\textrm{at}\,1\,\mathrm{Hz}$ and zero white noise. Wavelet edge detection performs well in this situation, achieving a score of $F=0.99$ over the full length of the signal. Threshold edge detection is significantly affected by $1/f$ noise, reaching only $F=0.79$. This low score is easily understood: large, low-frequency components of the noise cause the signal to drift with respect to the threshold level, resulting in false positives and missed events. In contrast, the same noise does little to distort the sharp edges in the data and these are easily identified in the wavelet transform.

The signal in Fig.~\ref{fig2}(d) was generated with zero $1/f$ noise and white noise of RMS amplitude $A_W = 0.008e\,/\mathrm{\sqrt{Hz}}$ (a peak-to-peak noise amplitude of $1.01e$ in this bandwidth). Over the full length of the signal, the wavelet edge detection has a score of $F=0.88$ and the threshold edge detection has $F=0.40$. The low score for thresholding is largely due to false positive detections, which is unsurprising given the low signal-to-noise ratio. The wavelet edge detection is still able to identify events because edges produce a distinct profile in the wavelet transform that is not reproduced by white noise: a local maximum that extends over a large range of scales.

\begin{figure}
\centering
\includegraphics{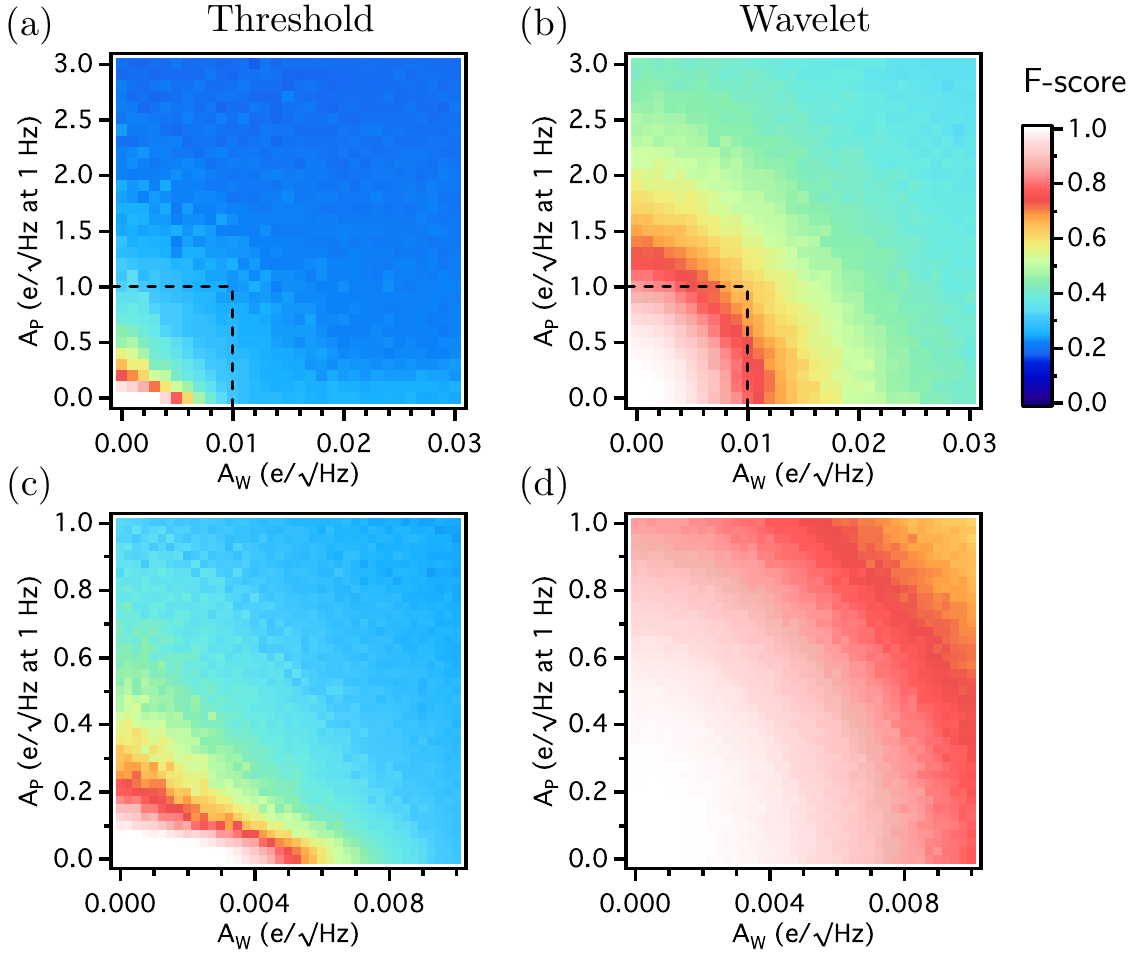}
\caption{Averaged F-score (see definition in main text) for wavelet edge detection (b),(d) and threshold edge detection (a),(c) as a function of white noise amplitude $A_W$ and $1/f$ noise amplitude $A_P$. (c) and (d) show detail within the black dashed boxes in (a) and (b) respectively. A score of $F=1$ corresponds to perfect detection. A score of $F=2/3$ corresponds, for example, to half the edges being detected with no false positives, or all edges being detected but with an equal number of false positives. The wavelet edge detection performs as well as threshold edge detection or better at all noise levels, and is significantly more robust against $1/f$ noise.\label{fig3}}
\end{figure}

The two detection techniques were compared over a range of noise amplitudes. Fig.~\ref{fig3}(a-d) shows F-scores for each value of $(A_W, A_P)$, which were found by averaging over 10 signals that were generated by adding random noise to the same ideal sensor output. Each signal is $2^{12}$ data points ($\approx 2\,\mathrm{s}$) long, and each plot in Fig.~\ref{fig3} is the result of analyzing $\sim 10,000$ signals. The results show that the wavelet edge detection matches or outperforms thresholding at all noise levels. From Fig.~\ref{fig3}(c) and (d), it is clear that the wavelet approach is significantly more tolerant to $1/f$ noise in the signal and, by extension, low-frequency noise in general.

There are several improvements that may be made to the wavelet edge detection algorithm. Typically, the slowest part of the analysis is tracking edges in the wavelet transform. The analysis time can be significantly reduced by choosing only to track maxima/minima in the transform once they exceed a certain weight. This ensures that large numbers of short, weak tracks, such as those produced by white noise, are discarded. The rate of false positives may be reduced by including a prior knowledge of the detector sensitivity: edges can be rejected if the signal level does not change by the expected value in the vicinity of the edge. It should be noted that there are also many ways to improve the thresholding technique presented here, although at the cost of adding additional parameters that must be tuned; for example, local averaging to compensate for low-frequency signal drift \cite{Lu2003}, or using a Schmitt trigger to select events \cite{Lambert2013}. In general, these filters will make threshold edge detection more similar to the wavelet approach, which can be thought of as a filter with an excellent selectivity to sharp edges.

\subsection{Comparison between wavelet and threshold detection using experimtanally measured signals}\label{sec:real_signals}

\begin{figure}
\centering
\includegraphics{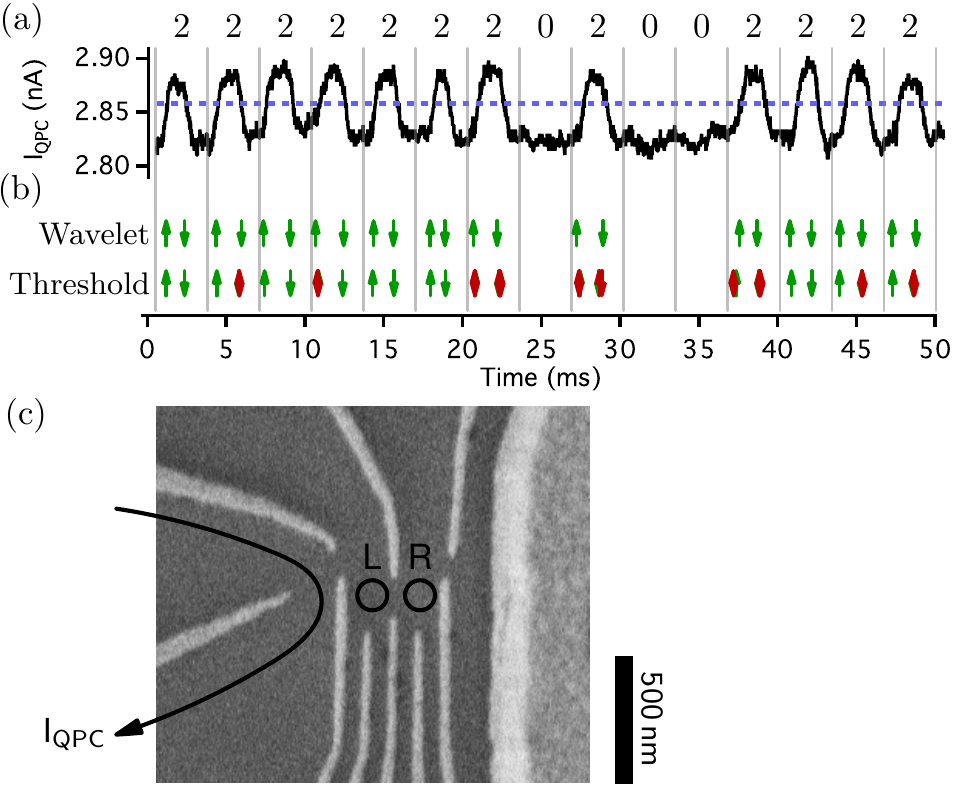}
\caption{Example charge sensor data from a quantum point-contact adjacent to a Si/SiGe double quantum dot \cite{Prance2012}. (a) Measured current $I_{QPC}$ through the charge sensor as the quantum dots are driven by a $300\,\mathrm{Hz}$ square wave. Either two or zero charging events are expected during each $\approx 3.3\,\mathrm{ms}$ pulse period. The periods are separated by vertical grey lines, and the observed number of events in each period are shown above the data. Pairs of events are visible in the signal as peaks (one upwards transition and one downwards). (b) Positions and directions of edges detected in the signal by the thresholding and wavelet algorithms. Wavelet detection produced no errors in this section of signal. Thresholding detection produced several errors, indicated by red arrows. In this section of the signal, the majority of these errors are multiple detections of the same edge due to high frequency signal noise.\label{fig4}}
\end{figure}

To demonstrate the benefits of wavelet edge detection we apply the technique to experimental data obtained from a point-contact charge sensor adjacent to a Si/SiGe double quantum dot. Fig.~\ref{fig4} shows an SEM image of the device. The charge sensor signal $I_{QPC}$ has two preferred values, corresponding to an additional electron being in either the left or right quantum dot. Further details of the device and the measurement setup can be found in Prance et al. \cite{Prance2012}. Importantly for this work, events in the charge sensor signal are known to occur at certain times, as discussed below, which makes it possible to independently estimate the accuracy of the edge detection. Based on this \emph{a priori} knowledge of the signal, we can compare the accuracy of the wavelet and thresholding algorithms using real data.

An example signal from the Si/SiGe point-contact charge sensor is shown in Fig.~\ref{fig4}. In contrast to the data used in section 3.1, the times at which edges occur in this signal are not random. During the measurement the double dot was driven by a $300\,\mathrm{Hz}$ square wave. Any charging events in the signal are aligned to this pulse with a high probability and there are only two expected outcomes within each pulse period: either no edges or an up-down pair of edges will be present. All other outcomes are known to be unlikely due to the behaviour of the double quantum dot \cite{Prance2012}. 

We use wavelet and threshold edge detection to find the number of edges during each pulse period. We estimate the error of each edge detection method to be the percentage of pulse periods that are found to contain an unexpected number of edges (i.e. not zero or two). Table~\ref{tbl1} shows the optimised errors using wavelet and thresholding edge detection as described in sections 2 and 3.1. The values in Table~\ref{tbl1} are based on an analysis of $120$ signals, each consisting of $10^5$ data points  (a total of $\approx 72,000$ pulse periods) sampled at a rate of $50\,\mathrm{kHz}$. For this dataset the wavelet detection is found to significantly outperform the simple thresholding algorithm. (The wavelet error is $\approx 2\%$ while the thresholding error is $\approx 50\%$).

\begin{table}
\centering
\begin{tabular}{l|c|c|c}
Edge detection method & Unexpected events \\
\hline
\hline
Wavelet & $2.26\%$ \\
Wavelet with $10\,\mathrm{kHz}$ filter & $2.24\%$ \\
\hline
Simple threshold & $50.2\%$ \\
Threshold with drift correction (d.c.) & $48.7\%$ \\
Threshold with $10\,\mathrm{kHz}$ filter & $13.7\%$ \\
Threshold with $10\,\mathrm{kHz}$ filter and d.c. & $12.5\%$ \\
\end{tabular}
\caption{Estimate of the characteristic error for several edge detection algorithms when applied to experimental charge sensor data. The error is estimated by comparing the detected edges to the expected characteristics of the signal (see Fig.~\ref{fig4}). For each result the free parameters of the corresponding algorithm were optimised to minimise the error. \label{tbl1}}
\end{table}

The performance of the thresholding algorithm can be improved at the cost of reduced bandwidth, extra free parameters, and the need for additional post-processing. If the threshold is defined relative to the average value of the signal, the effect of low frequency signal drift can be reduced. This drift correction decreases the error marginally to $\approx 49\%$. A greater improvement is found by applying a low-pass filter. When the signal bandwidth is reduced to $10\,\mathrm{kHz}$, the error is reduced to $\approx 14\%$. Combining this low-pass filter with drift correction gives an error of $\approx 13\%$. The low pass filter has negligible effect on the wavelet edge detection, which already filters the signal at multiple scales as part of its operation. While it is possible to achieve reasonable results using thresholding on this data set, the resulting algorithm includes one additional free parameter, the filter cut-off frequency, and needs knowledge of the mean signal level, which requires averaging over long times. Furthermore, while the $10\,\mathrm{kHz}$ filter reduces the error in this particular data set, it limits the detection bandwidth in general. By contrast, the wavelet edge detection performs well with just one free parameter and no additional filtering.

\section{Conclusions}

We have compared the performance of a wavelet edge detection algorithm to a thresholding methods using simulated signals that mimic the output of a QPC charge sensor and measured signals from a Si/SiGe double quantum dot device. When only a single free parameter is allowed in the detection algorithms, wavelet edge detection is found to outperform thresholding over all noise profiles in the simulated signals and is particularly robust against $1/f$ noise. Wavelet detection is also found to perform well on real data, even when compared to a more complex thresholding algorithm with additional free parameters. A significant practical advantage is that, once optimized, the wavelet edge detection can be used in the presence of low-frequency noise and signal drift without adjustment. This means that the analysis may be performed while the experiment is running, which is crucial for certain qubit applications such as error correction. Furthermore, the excellent sensitivity of wavelet edge detection can be used to increase the fidelity and potentially improve the bandwidth of qubit readout based on a single electron charge sensor.

\ack

This research was supported in part by the U.S. Army Research Office (W911NF-08-1-0482, W911NF-12-1-0607). Development and maintenance of the growth facilities used for fabricating samples is supported by DOE (DE-FG02-03ER46028). This research utilized NSF-supported shared facilities at the University of Wisconsin-Madison.

\section*{References}

\providecommand{\newblock}{}


\begin{thebibliography}{10}
\expandafter\ifx\csname url\endcsname\relax
  \def\url#1{{\tt #1}}\fi
\expandafter\ifx\csname urlprefix\endcsname\relax\def\urlprefix{URL }\fi
\providecommand{\eprint}[2][]{\url{#2}}

\bibitem{Field1993}
Field M, Smith C~G, Pepper M, Ritchie D~A, Frost J~E~F, Jones G~A~C and Hasko
  D~G 1993 {\em Phys. Rev. Lett.\/} {\bf 70} 1311--1314

\bibitem{Elzerman2003}
Elzerman J~M, Hanson R, Greidanus J~S, Willems~van Beveren L~H, De~Franceschi
  S, Vandersypen L~M~K, Tarucha S and Kouwenhoven L~P 2003 {\em Phys. Rev. B\/}
  {\bf 67} 161308

\bibitem{Petta2004}
Petta J~R, Johnson A~C, Marcus C~M, Hanson M~P and Gossard A~C 2004 {\em Phys.
  Rev. Lett.\/} {\bf 93} 186802

\bibitem{Astafiev2004}
Astafiev O, Pashkin Y~A, Yamamoto T, Nakamura Y and Tsai J~S 2004 {\em Phys.
  Rev. B\/} {\bf 69}(18) 180507

\bibitem{Shi2013}
Shi Z, Simmons C~B, Ward D~R, Prance J~R, Mohr R~T, Koh T~S, Gamble J~K, Wu X,
  Savage D~E, Lagally M~G, Friesen M, Coppersmith S~N and Eriksson M~A 2013
  {\em Phys. Rev. B\/} {\bf 88}(7) 075416

\bibitem{Elzerman2004}
Elzerman J~M, Hanson R, Willems~van Beveren L~H, Witkamp B, Vandersypen L~M~K
  and Kouwenhoven L~P 2004 {\em Nature\/} {\bf 430} 431--435

\bibitem{Hanson2005}
Hanson R, van Beveren L~H~W, Vink I~T, Elzerman J~M, Naber W~J~M, Koppens
  F~H~L, Kouwenhoven L~P and Vandersypen L~M~K 2005 {\em Phys. Rev. Lett.\/}
  {\bf 94} 196802

\bibitem{Johnson2005a}
Johnson A~C, Petta J~R, Taylor J~M, Yacoby A, Lukin M~D, Marcus C~M, Hanson M~P
  and Gossard A~C 2005 {\em Nature\/} {\bf 435} 925--928

\bibitem{Petta2005}
Petta J, Johnson A, Taylor J, Laird E, Yacoby A, Lukin M, Marcus C, Hanson M
  and Gossard A 2005 {\em Science\/} {\bf 309} 2180--2184

\bibitem{Shi2012}
Shi Z, Simmons C~B, Prance J~R, Gamble J~K, Koh T~S, Shim Y~P, Hu X, Savage
  D~E, Lagally M~G, Eriksson M~A, Friesen M and Coppersmith S~N 2012 {\em Phys.
  Rev. Lett.\/} {\bf 108}(14) 140503

\bibitem{Kim2014}
Kim D, Shi Z, Simmons C~B, Ward D~R, Prance J~R, Koh T~S, Gamble J~K, Savage
  D~E, Lagally M~G, Friesen M, Coppersmith S~N and Eriksson M~A 2014 {\em
  Nature\/} {\bf 511} 70--74

\bibitem{Ladd2010}
Ladd T~D, Jelezko F, Laflamme R, Nakamura Y, Monroe C and O'Brien J~L 2010 {\em
  Nature\/} {\bf 464} 45--53

\bibitem{Nowack2011}
Nowack K~C, Shafiei M, Laforest M, Prawiroatmodjo G~E~D~K, Schreiber L~R,
  Reichl C, Wegscheider W and Vandersypen L~M~K 2011 {\em Science\/} {\bf 333}
  1269--1272

\bibitem{Shulman2012}
Shulman M~D, Dial O~E, Harvey S~P, Bluhm H, Umansky V and Yacoby A 2012 {\em
  Science\/} {\bf 336} 202--205

\bibitem{Barthel2012}
Barthel C, Medford J, Bluhm H, Yacoby A, Marcus C~M, Hanson M~P and Gossard A~C
  2012 {\em Phys. Rev. B\/} {\bf 85}(3) 035306

\bibitem{Fink2013}
Fink T and Bluhm H 2013 {\em Phys. Rev. Lett.\/} {\bf 110}(1) 010403

\bibitem{Lu2003}
Lu W, Ji Z, Pfeiffer L, West K~W and Rimberg A~J 2003 {\em Nature\/} {\bf 423}
  422--425 ISSN 0028-0836

\bibitem{Fujisawa2004a}
Fujisawa T, Hayashi T, Hirayama Y, Cheong H~D and Jeong Y~H 2004 {\em Appl.
  Phys. Lett.\/} {\bf 84} 2343--2345

\bibitem{Vandersypen2004}
Vandersypen L~M~K, Elzerman J~M, Schouten R~N, van Beveren L~H~W, Hanson R and
  Kouwenhoven L~P 2004 {\em Appl. Phys. Lett.\/} {\bf 85} 4394--4396

\bibitem{Schleser2004}
Schleser R, Ruh E, Ihn T, Ensslin K, Driscoll D~C and Gossard A~C 2004 {\em
  Appl. Phys. Lett.\/} {\bf 85} 2005--2007

\bibitem{Meunier2006}
Meunier T, Vink I~T, Willems~van Beveren L~H, Koppens F~H~L, Tranitz H~P,
  Wegscheider W, Kouwenhoven L~P and Vandersypen L~M~K 2006 {\em Phys. Rev.
  B\/} {\bf 74}(19) 195303

\bibitem{Gustavsson2006a}
Gustavsson S, Leturcq R, Simovic B, Schleser R, Ihn T, Studerus P, Ensslin K,
  Driscoll D~C and Gossard A~C 2006 {\em Phys. Rev. Lett.\/} {\bf 96} 076605
  (pages~4)

\bibitem{Reilly2007}
Reilly D~J, Marcus C~M, Hanson M~P and Gossard A~C 2007 {\em Appl. Phys.
  Lett.\/} {\bf 91} 162101 (pages~3)

\bibitem{Barthel2009}
Barthel C, Reilly D~J, Marcus C~M, Hanson M~P and Gossard A~C 2009 {\em Phys.
  Rev. Lett.\/} {\bf 103}(16) 160503

\bibitem{Gustavsson2009}
Gustavsson S, Leturcq R, Studer M, Shorubalko I, Ihn T, Ensslin K, Driscoll D
  and Gossard A 2009 {\em Surf. Sci. Rep.\/} {\bf 64} 191 -- 232 ISSN 0167-5729

\bibitem{Mason2010}
Mason J, Gaudreau L, Studenikin S, Kam A, Djurkovic B, Sachrajda A and Kycia J
  2010 {\em Physica E\/} {\bf 42} 813 -- 816 ISSN 1386-9477

\bibitem{Gustavsson2008a}
Gustavsson S, Shorubalko I, Leturcq R, Sch\"{o}n S and Ensslin K 2008 {\em
  Appl. Phys. Lett.\/} {\bf 92} 152101 (pages~3)

\bibitem{Gotz2008}
Gotz G, Steele G~A, Vos W~J and Kouwenhoven L~P 2008 {\em Nano Letters\/} {\bf
  8} 4039--4042 pMID: 18928322

\bibitem{Guttinger2011}
G\"uttinger J, Seif J, Stampfer C, Capelli A, Ensslin K and Ihn T 2011 {\em
  Phys. Rev. B\/} {\bf 83} 165445

\bibitem{Morello2010}
Morello A, Pla J~J, Zwanenburg F~A, Chan K~W, Tan K~Y, Huebl H,
  M{\"o}tt{\"o}nen M, Nugroho C~D, Yang C, van Donkelaar J~A, Alves A~D~C,
  {Jamieson} D~N, Escott C~C, Hollenberg L~C~L, Clark R~G and Dzurak A~S 2010
  {\em Nature\/} {\bf 467} 687--691

\bibitem{Thalakulam2011}
Thalakulam M, Simmons C~B, Van~Bael B~J, Rosemeyer B~M, Savage D~E, Lagally
  M~G, Friesen M, Coppersmith S~N and Eriksson M~A 2011 {\em Phys. Rev. B\/}
  {\bf 84}(4) 045307

\bibitem{Simmons2011}
Simmons C~B, Prance J~R, Van~Bael B~J, Koh T~S, Shi Z, Savage D~E, Lagally M~G,
  Joynt R, Friesen M, Coppersmith S~N and Eriksson M~A 2011 {\em Phys. Rev.
  Lett.\/} {\bf 106} 156804

\bibitem{Mahapatra2011}
Mahapatra S, Buch H and Simmons M~Y 2011 {\em Nano Letters\/} {\bf 11}
  4376--4381

\bibitem{Prance2012}
Prance J~R, Shi Z, Simmons C~B, Savage D~E, Lagally M~G, Schreiber L~R,
  Vandersypen L~M~K, Friesen M, Joynt R, Coppersmith S~N and Eriksson M~A 2012
  {\em Phys. Rev. Lett.\/} {\bf 108}(4) 046808

\bibitem{Yuan2012}
Yuan M, Yang Z, Savage D~E, Lagally M~G, Eriksson M~A and Rimberg A~J 2012 {\em
  Appl. Phys. Lett.\/} {\bf 101} 142103 (pages~3)

\bibitem{Schoelkopf1998}
Schoelkopf R~J, Wahlgren P, Kozhevnikov A~A, Delsing P and Prober D~E 1998 {\em
  Science\/} {\bf 280} 1238--1242

\bibitem{Jung2004}
Jung S~W, Fujisawa T, Hirayama Y and Jeong Y~H 2004 {\em Appl. Phys. Lett.\/}
  {\bf 85} 768--770

\bibitem{ENCODE2007}
{The ENCODE Project Consortium} 2007 {\em Nature\/} {\bf 447} 799--816

\bibitem{JPEG2000}
{Joint Photographic Experts Group} {JPEG 2000 standard}
  \urlprefix\url{http://www.jpeg.org/jpeg2000}

\bibitem{Bradley1993}
Bradley J~N, Brislawn C~M and Hopper T 1993 {\em Proc. SPIE\/} {\bf 1961}
  293--304

\bibitem{Mallat2009}
Mallat S 2009 {\em A Wavelet Tour of Signal Processing\/} 3rd ed (Elsevier
  Academic Press)

\bibitem{Lambert2013}
Lambert N~J, Edwards M, Esmail A~A, Pollock F~A, Barrett S~D, Lovett B~W and
  Ferguson A~J 2013 {\em ArXiv e-prints\/} (\textit{Preprint}
  \eprint{1304.5117})

\bibitem{Canny1986}
Canny J 1986 {\em Pattern Analysis and Machine Intelligence, IEEE Transactions
  on\/} {\bf PAMI-8} 679--698 ISSN 0162-8828

\bibitem{WaveLab2013}
The function `RWT' in WaveLab850 is used to calculate the wavelet transform.
  The function `MM\_RWT' is used to find the location of local maxima in the
  modulus of the transform.
  \urlprefix\url{http://www-stat.stanford.edu/~wavelab}

\bibitem{Rijsbergen1979}
van Rijsbergen C~J 1979 {\em Information Retrieval\/} 2nd ed (Butterworths)

\end{thebibliography}
\end{document}